\documentclass[preprint,showpacs,amsmath,amssymb,aps,prd,nofootinbib]{revtex4}
\usepackage{epsfig,graphicx,color,appendix}
\begin{document}
\begin{flushright}
IBS-CTPU-17-20
\end{flushright}
\title{\mbox{}\\[10pt]
A necessary condition for sphaleron process\\ in the presence of anomalous $U(1)$ symmetry}

\author{Y. H. Ahn}
\affiliation{Center for Theoretical Physics of the Universe, Institute for Basic Science (IBS), Daejeon, 34051, Korea}
\email{yhahn@ibs.re.kr}


\begin{abstract}
\noindent We argue that, in the presence of anomalous $U(1)$ symmetries, the invariance of Lagrangian including the standard model (SM) under the axionic shift symmetries requires a necessary condition $\sum_i\delta^{\rm GS}_{i}\leq2N_f$, where $N_f$ stands for the number of families in the SM and $\delta^{\rm GS}_i$ (Green-Schwarz parameter) characterizes the coupling of the anomalous $U(1)$ gauge boson to the corresponding axion. 
In turn, we show that in order for the usual $B+L$ violating sphaleron process to be valid a necessary condition $\sum_i\delta^{\rm GS}_{i}<2N_f$ is required, where $B(L)$ stands for the baryon(lepton) number.
\end{abstract}
\maketitle %
Symmetries play an important role in physics in general and in quantum field theory in particular. A symmetry of the classical action is a transformation of the fields that leaves the action invariant. Anomaly breaks any classical symmetry of the Lagrangian at the quantum level. So, all local gauge theories must be free of anomalies.
The standard model (SM) as an effective gauge theory has been successful in describing phenomena until now, but it suffers from theoretical arguments (inclusion of gravity in gauge theory, instability of the Higgs potential, the SM fermion mass hierarchies and their mixing patterns with the CP violating phases, the strong CP problem~\cite{Peccei:1977hh}, etc) and cosmological evidences (dark matter, inflation, cosmological constant, etc). It is widely believed that the SM should be extended to a more fundamental underlying theory.

Once gauged $U(1)_{X_i}$ symmetries are introduced in the SM to solve aforementioned theoretical arguments or cosmological evidences, there could be anomalous current $J^{X_i}_\mu$ coupling to their gauge bosons, that is, $\partial_{\mu}J^{\mu}_{X_i}=\frac{\delta^{\rm GS}_i}{16\pi^2}{\rm Tr}(Q^{\mu\nu}\tilde{Q}_{\mu\nu})$\,\cite{ABJ}, where $\tilde{Q}_{\mu\nu}\equiv\frac{1}{2}\epsilon^{\mu\nu\rho\sigma}Q_{\rho\sigma}$ is the dual of the field strength tensor of the $\{SU(3)_C, SU(2)_L, U(1)_Y\}$ field with $Q=\{G, W, H\}$, respectively; the Green-Schwarz parameter $\delta^{\rm GS}_i$ characterizes the coupling of the anomalous gauge boson to the axion\,\cite{Ahn:2016typ, Ahn:2016hbn}. In this paper we only focus on the study of $SU(2)$ instantonic configuration; see Ref.\,\cite{Ahn:2016typ} for $SU(3)_C$ instantonic configuration.
The gauged $U(1)_{X_i}$ symmetries are broken, and after decoupling their corresponding gauge bosons which are assumed to be very heavy, only the SM gauge group remains at low energy.
We consider the case that anomalies in local gauge theory are removed by the Green-Schwarz mechanism\,\cite{GS}, while leaving behind the ``anomalous global" $U(1)_{X_i}$ symmetries with the low energy effective Lagrangian\,\cite{Ahn:2016hbn}
\begin{eqnarray}
 {\cal L}\supset{\cal L}_{\rm SM}+\sum_i\frac{\delta^{Q}_i}{16\pi^2}\frac{A_i}{f_{a_i}}{\rm Tr}(Q^{\mu\nu}\tilde{Q}_{\mu\nu})\,,
 \label{axi_lag}
\end{eqnarray}
where  $A_i$ is the axion field with its decay constant $f_{a_i}$ ($i=1,2,..$), and $\delta^{Q}_i$ stand for the coefficients of the mixed  $U(1)_{X_i}$-$[SU(2)_L]^2$, and $U(1)_{X_i}$-$[U(1)_Y]^2$ anomalies.  Under the axionic shift symmetry
\begin{eqnarray}
 \frac{A_i}{f_{a_i}}\rightarrow\frac{A_i}{f_{a_i}}+\frac{\delta^{\rm GS}_i}{\delta^Q_i}\xi\,,
\label{shif}
\end{eqnarray}
the second term in the right-hand side in Eq.\,(\ref{axi_lag}) transforms
\begin{eqnarray}
 \frac{\delta^Q_i}{\delta^{\rm GS}_{i}}\partial_{\mu}J^\mu_{Qi}\frac{A_i}{f_{a_i}}\rightarrow-\frac{\delta^Q_i}{\delta^{\rm GS}_{i}}J^\mu_{Qi}\partial_\mu\frac{A_i}{f_{a_i}}+\xi\partial_{\mu}J^\mu_{Qi}\,.
\label{lag_shif}
\end{eqnarray}
As a total divergence we can express the last term, $\partial_{\mu}J^\mu_{Qi}\equiv\partial_{\mu}J^\mu_{Xi}$, in Eq.\,(\ref{lag_shif}):
\begin{eqnarray}
 \partial_\mu\left(J^\mu_{Wi}+J^\mu_{Hi}\right)=\delta^{\rm GS}_{i}\,\partial_\mu\left(K^\mu+k^\mu\right)\,,
 \label{kk}
\end{eqnarray}
where
\begin{eqnarray}
 &&K^\mu=\frac{\epsilon^{\mu\nu\rho\sigma}}{32\pi^2}2W^a_\nu\left(\partial_{\rho}W^a_\sigma+\frac{\epsilon^{abc}}{3}W^b_{\rho}W^c_{\sigma}\right)\,,\nonumber\\
 &&k^\mu=\frac{\epsilon^{\mu\nu\rho\sigma}}{32\pi^2}B_{\nu}B_{\rho\sigma}\,.
 \label{kk0}
\end{eqnarray}
Here $W^a_\mu$ ($a=1-3$) and $B_\mu$ are weak and hypercharge gauge fields, respectively, and their corresponding gauge couplings are absorbed.
Since, in a $U(1)$ gauge theory, the resulting surface term in the action at infinite would vanish for finite energy configurations, the term $\partial_{\mu}J^{\mu}_{Hi}$ in Eq.\,(\ref{kk}) corresponding to $\partial_{\mu}k^\mu$ goes to zero. On the other hand, for the non-Abelian part $K^\mu$ the integral over all space-time with vanishing boundary conditions does not necessarily vanish. Thus, under the axionic shift symmetry in Eq.\,(\ref{lag_shif}) the Lagrangian\,(\ref{axi_lag}) is not invariant due to the last term in Eq.\,(\ref{lag_shif}), $\partial_{\mu}J^{\mu}_{Wi}$. How to make it invariant under the axionic shift symmetry ?

In this paper we argue that $\partial_{\mu}J^{\mu}_{W}=\sum_i\partial_{\mu}J^{\mu}_{Wi}$ corresponding  to the $SU(2)$ weak vacuum structure could be removed by the $B+L$ transformation, where $B(L)$ stands for the baryon(lepton) number. Along this line, we will show a condition for $B+L$ violating sphaleron process\,\cite{Kuzmin:1985mm} .

In the SM the chiral nature of the weak interactions makes the baryon- and lepton number anomalies. The $B$- and $L$ currents defined by $J^{\mu}_B=\frac{1}{3}\sum(\bar{q}_L\gamma^{\mu}q_L-\bar{u}_R\gamma^{\mu}u_R-\bar{d}_R\gamma^{\mu}d_R)$ and $J^{\mu}_L=\sum(\bar{\ell}_L\gamma^{\mu}\ell_L-\bar{\ell}_R\gamma^{\mu}\ell_R)$, where the sum runs over families and colors (for the quarks), are broken by $U(1)_B$ and $U(1)_L$ anomalies, respectively:
\begin{eqnarray}
 \partial_{\mu}J^{\mu}_B=\partial_{\mu}J^{\mu}_L&=&-\frac{N_f}{16\pi^2}{\rm Tr}(W_{\mu\nu}\tilde{W}^{\mu\nu})\nonumber\\
 &=&-N_f\,\partial_{\mu}K^{\mu}\,,
 \label{bl}
\end{eqnarray}
where $N_f$ is the number of families, and similar to $\partial_{\mu}J^\mu_{Hi}=\delta^{\rm GS}_{i}\partial_{\mu}k^\mu$ the Abelian part was neglected. 
Immediately, Eq.\,(\ref{bl}) implies that
\begin{eqnarray}
 \partial_\mu\left(J^{\mu}_{B}-J^{\mu}_{L}\right)&=&0\,,\nonumber\\
 \partial_\mu\left(J^{\mu}_{B}+J^{\mu}_{L}\right)&=&-\frac{N_f}{8\pi^2}{\rm Tr}(W_{\mu\nu}\tilde{W}^{\mu\nu})\,,
  \label{bpl}
\end{eqnarray}
that is, $B-L$ is strictly conserved, while $B+L$ is violated due to vacuum structure of $SU(2)$ non-Abelian gauge theory. 

Meanwhile, in the presence of anomalous $U(1)_{X_i}$ the Lagrangian (\ref{axi_lag}) clearly transforms under the shift symmetries in Eq.\,(\ref{shif}) as
\begin{eqnarray}
 {\cal L}\supset{\cal L}_{\rm SM}+\sum_i\frac{\delta^{Q}_i}{16\pi^2}\frac{A_i}{f_{a_i}}{\rm Tr}(Q^{\mu\nu}\tilde{Q}_{\mu\nu})+\xi\,\partial_{\mu}J^{\mu}_{W},
 \label{axi_lag1}
\end{eqnarray}
where $\partial_\mu J^{\mu}_{W}\equiv\sum_i\delta^{\rm GS}_{i}\,\partial_{\mu}K^{\mu}$. So, in order for the Lagrangian\,(\ref{axi_lag}) to be invariant under the axionic shift symmetries in Eq.\,(\ref{shif}) the $SU(2)$ weak vacuum structure $\partial_{\mu}J^{\mu}_{W}$ should be removed through the $B+L$ transformation
\begin{eqnarray}
 \partial_\mu\left(J^{\mu}_{W}+J^{\mu}_{B+L}\right)&=&\frac{\sum_i\delta^{\rm GS}_{i}-2N_f}{16\pi^2}{\rm Tr}(W_{\mu\nu}\tilde{W}^{\mu\nu})\,,
\label{gau_inv}
\end{eqnarray}
with the relation $\partial_\mu\left(J^{\mu}_{B}-J^{\mu}_{L}\right)=0$ kept in ${\cal L}_{\rm SM}$. Then, the invariance of the Lagrangian (\ref{axi_lag1}) under the axionic shift symmetries in Eq.\,(\ref{shif}) requires a necessary condition
\begin{eqnarray}
 \sum_i\delta^{\rm GS}_{i}\leq2N_f\,.
\label{gau_inv1}
\end{eqnarray}
In other words, the anomalous current $J^{\mu}_{W}$ coupling to the $U(1)_{X_i}$ gauge bosons $A^\mu_i$ could be removed by the anomalous current $J^{\mu}_{B+L}\equiv J^{\mu}_{B}+J^{\mu}_{L}$ with the above necessary condition. In turn, for the condition $\sum_i\delta^{\rm GS}_{i}=2N_f$ both directions $B\pm L$ could be anomaly-free, leading to no $B+L$ violating sphaleron process; while for $\sum_i\delta^{\rm GS}_{i}<2N_f$ the $B-L$ is conserved but $B+L$ is not, leading to the usual $B+L$ violating sphaleron process.
At finite temperature there exists non-trivial static topological soliton configuration, called ``sphalerons"; thermal fluctuations in the $SU(2)$ gauge field $W^a_\mu$ and symmetry-breaking Higgs field $H$ can cause transitions that proceed over the potential barrier between two different $\theta$-vacua, thereby resulting in baryon and lepton number violation\,\cite{Kuzmin:1985mm}. Actually, the sphaleron configurations are spatially large. So, at high temperature the Boltzmann barrier becomes smaller (or vanishing) and the transition that connect different vacua easily takes place.

We conclude that in the presence of anomalous $U(1)_{X_i}$ 
in order for the usual $B+L$ violating sphaleron process to be valid a necessary condition $\sum_i\delta^{\rm GS}_{i}<2N_f$ is required.

\acknowledgments{This work was supported by IBS under the project code, IBS-R018-D1. We would like to give thanks to Thomas Flacke for useful discussion.
}


\end{document}